\documentclass[letterpaper,twocolumn,10pt,conference]{IEEEtran}
\usepackage[utf8]{inputenc}

\usepackage[cmex10]{amsmath} % cmex10 proposed by IEEEtran_HOWTO
\usepackage{amssymb,amsthm,bm}
\usepackage{mathtools}
\usepackage{bm}
\usepackage{url}
\usepackage{cite}
\usepackage{tikz}
\usetikzlibrary{calc,positioning,fit,shapes,backgrounds}
\usepackage{pgfplots}
\usepgfplotslibrary{fillbetween}
\pgfplotsset{compat=1.17}

\usepackage{subcaption}
\usepackage{calc}

\newtheorem{theorem}{Theorem}
\newtheorem{lemma}{Lemma}

\newtheorem{definition}{Definition}

\newcommand{\given}{\, |\,}

\usepackage{bbm}

\newcommand{\EE}{\mathbb{E}}
\newcommand{\RR}{\mathbb{R}}
\newcommand{\barR}{\bar{\mathbb{R}}}
\DeclareMathOperator{\ri}{ri}
\DeclareMathOperator{\dom}{dom}

\newcommand{\pWW}{\sffamily NUNP}
\newcommand{\pWR}{\sffamily NURP}
\newcommand{\pRW}{\sffamily RUNP}
\newcommand{\pRR}{\sffamily RURP}

\title{Robust Optimization for Local Differential Privacy}
\author{
\IEEEauthorblockN{Jasper Goseling}
\IEEEauthorblockA{
 University of Twente, \\
 the Netherlands
}
\and 
\IEEEauthorblockN{Milan Lopuhaä-Zwakenberg}
\IEEEauthorblockA{
 University of Twente, \\
 the Netherlands
}
}
\date{June 2022}

\begin{document}

\maketitle

\begin{abstract}
We consider the setting of publishing data without leaking sensitive information. We do so in the framework of Robust Local Differential Privacy (RLDP). This ensures privacy for all distributions of the data in an uncertainty set. We formulate the problem of finding the optimal data release protocol as a robust optimization problem. By deriving closed-form expressions for the duals of the constraints involved we obtain a convex optimization problem. We compare the performance of four possible optimization problems depending on whether or not we require robustness in i) utility and ii) privacy.  
\end{abstract}

%%%%%%%%%%%%%%%%%%%%%%%%%%%%%%%%%%%%%%%%%%%%%%%%%%%%%%%%%%%%%%%%%%%%%%%%%
%
%
%
%%%%%%%%%%%%%%%%%%%%%%%%%%%%%%%%%%%%%%%%%%%%%%%%%%%%%%%%%%%%%%%%%%%%%%%%%
\section{Introduction}
We consider a setting in which users have data $(S,U)$ with $U \in \mathbb{R}$. A user wants to publish $U$, but does not want to disclose information about their sensitive data $S$, which may be correlated with $U$. Therefore, the users release an obfuscated version $Y$ of $U$, such that $Y$ is as close to $U$ as possible, measured by mean squared distortion, without leaking too much information about $S$. This scenario and closely related ones have been studied in, for instance, ~\cite{rebollo2010t,kairouz2016extremal,makhdoumi2014information,salamatian2015managing,asoodeh2016information,kung2018compressive,ding2019submodularity,salamatian2020privacy,lopuhaa2021privacy,lopuhaa2021robust}.

To measure the information leakage about $S$ in $Y$ we use a form of Local Differential Privacy \cite{kasiviswanathan2011can} that was introduced in \cite{lopuhaa2020privacy}, which states that for each possible $s_1,s_2$ and $y$ the following should hold:
\begin{equation} \label{eq:privacy}
\mathbb{P}(Y = y|S = s_1) \leq \textrm{e}^{\varepsilon}\mathbb{P}(Y = y|S=s_2).
\end{equation}
This condition is less strict than regular LDP, reflecting the fact that only $S$, and not $U$ itself, needs to be protected. Note that \eqref{eq:privacy} depends on the joint probability distribution $P_{SU}$. A privacy protocol is given by the conditional probability distribution $P_{Y|SU}$, and given such a protocol, the distortion is measured by $\mathbb{E}(U-Y)^2$, which also depends on $P_{SU}$.

Thus both privacy and utility depend on the distribution of the data. However, a user may not know this distribution exactly and needs to estimate it. The odds ratio interpretation of differential privacy \cite{kifer2014pufferfish} tells us that an attacker with a better estimate may obtain more information about $S$ than the privacy protocol that is developed based on this estimate would indicate. Furthermore, a good utility under the estimated distribution may not imply a good utility under the actual distribution.

In this paper, therefore, we demand stronger, \emph{robust} privacy and utility guarantees. More concretely, we assume that there exists a publicly available dataset from which the users produce an estimate distribution $\hat{P}_{SU}$, and a set $\mathcal{F}$ of probability distributions that do not differ significantly from $\hat{P}_{S,U}$ (for a chosen significance level). Then robustness can be incorporated in the following two ways:
\begin{enumerate}
    \item We demand robust privacy by requiring \eqref{eq:privacy} to hold for all $P_{SU} \in \mathcal{F}$;
    \item We obtain robust utility by minimising $\max_{P_{SU} \in \mathcal{F}} \mathbb{E} (U-Y)^2$.
\end{enumerate}
Robustness allows us to guarantee privacy and utility for all probability distributions one can reasonably expect, without sacrificing too much utility to account for unlikely distributions. It is important to note that: i) robustness in utility is w.r.t.\ our own uncertainty about $P_{S,U}$ and ii) robustness in privacy is w.r.t.\ our uncertainty about current knowledge by an attacker. The robustness of privacy and utility can be incorporated independently, leaving us with four possible optimization problems depending on for which of the two we want robust guarantees.

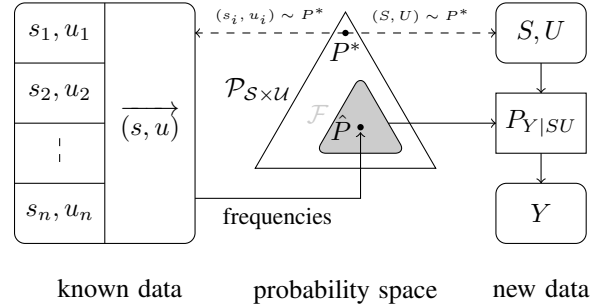
\begin{figure}
    \centering
\begin{tikzpicture}[scale=0.4]

\draw[rounded corners] (-7,0.5) -- (-13,0.5) -- (-13,-7.5) -- (-7,-7.5) -- cycle;
\draw[-] (-10,0.5) -- (-10,-7.5);
\draw[-] (-10,-1.5) -- (-13,-1.5);
\draw[-] (-10,-3.5) -- (-13,-3.5);
\draw[-] (-10,-5.5) -- (-13,-5.5);
\draw (-8.5,-3.5) node {$\overrightarrow{(s,u)}$};
\draw (-11.5,-0.5) node {$s_1,u_1$};
\draw (-11.5,-2.5) node {$s_2,u_2$};
\draw (-11.5,-6.5) node {$s_n,u_n$};

\draw[-, dashed] (-11.5,-4) -- (-11.5,-5);

%\draw[-latex,decorate,decoration={snake}] (3,-1) --node[above]{$X_i \sim P^*$} (7,-1);
%\draw[-latex] (7,-5) --node[above]{frequencies} (3,-5);
\draw (-9.5,-9) node {known data};
\draw (-2,-9.1) node{probability space};
\draw (4.5,-9) node {new data};
\draw (1,-5) -- (-5,-5) --node[left]{$\mathcal{P}_{\mathcal{S}\times \mathcal{U}}$} (-2,0.1961524227) -- cycle;
\draw[rounded corners,text = black, draw = black, fill = black, fill opacity = 0.2, text opacity = 1] (0,-4.5) -- (-3,-4.5) --node[left]{$\mathcal{F}$} (-1.5,-1.90192378865) -- cycle;
\fill (-1.5,-3.63397459622) circle[radius=3pt] node[left]{$\hat{P}$};
\fill (-2,-0.5) circle[radius=3pt] node[below]{$P^*$};
\draw[->] (-7,-6) --node[below]{\small frequencies} (-1.5,-6) -- (-1.5,-3.8);
\draw[->,dashed] (-2,-0.5) --node[above]{\tiny $(s_i,u_i) \sim P^*$} (-7,-0.5);
\draw[->,dashed] (-2,-0.5) --node[above]{\tiny $(S,U) \sim P^*$} (3,-0.5);
\draw[rounded corners] (3,0.5) -- (6,0.5) -- (6,-1.5) -- (3,-1.5) -- cycle;
\draw (4.5,-0.5) node{$S,U$};
\draw (6,-2.5) -- (3,-2.5) -- (3,-4.5) -- (6,-4.5) -- cycle;
\draw (4.5,-3.5) node{$P_{Y|SU}$};
\draw[->] (-0.57735026919,-3.5) -- (3,-3.5);
\draw[rounded corners] (3,-5.5) -- (6,-5.5) -- (6,-7.5) -- (3,-7.5) -- cycle;
\draw (4.5,-6.5) node{$Y$};
\draw[->] (4.5,-1.5) -- (4.5,-2.5);
\draw[->] (4.5,-4.5) -- (4.5,-5.5);
\end{tikzpicture}
    \caption{Overview. For clarity we write $P^* = P^*_{SU}$ and $\hat{P} = \hat{P}_{SU}$. Since $\mathcal{F}$ is a confidence interval $P^*$ may or may not be an element of $\mathcal{F}$.}
    \label{fig:overview}
\end{figure}

In recent work~\cite{lopuhaa2021privacy,lopuhaa2021robust} we introduced the robust privacy framework that we also use here. An important difference is that in~\cite{lopuhaa2021privacy,lopuhaa2021robust} the utility measure is $\operatorname{I}(S, U; Y)$, leading to an optimization problem is not convex. Therefore, similar to~\cite{kairouz2016extremal}, the resulting techniques for analysis in~\cite{lopuhaa2021privacy,lopuhaa2021robust} are combinatorial in nature and heuristics are developed. In the current work the utility measure is the mean-squared error, leading to a non-robust optimization problem that is convex.

One of the main contributions of this paper is to provide convex formulations of the corresponding robust optimization problems. These formulations can be handled by standard convex optimization solvers in order to obtain privacy protocols that are provably optimal. Our numerical results demonstrate that: i) without including robustness on the privacy constraints the effective privacy guarantees that are obtained are very weak, ii) including robustness on privacy leads to a significant penalty on utility, and iii) including a robustness constraint on utility does not have a large impact if privacy robustness is already imposed.

Our techniques are rooted in robust optimization~\cite{ben2009robust, ben2015deriving}, an important aspect of which is to use Fenchel duality on constraints like~\eqref{eq:privacy} that need to hold for all distributions in $\mathcal{F}$. In~\cite{ben2013robust} and~\cite{ bertsimas2018data} duals are derived for constraints involving probabilities and uncertainty sets based on the $\chi^2$-test. A major difference in the current work, and a contribution on the technical side, is that~\eqref{eq:privacy} involves two conditional distributions from $\mathcal{F}$.

The structure of this paper is as follows. In Section~\ref{sec:model} we present the details of our model. We provide background on robust optimization in Section~\ref{sec:robustopt} and present our analytical results in Section~\ref{sec:results}. Insights obtained by numerical experiments are given Section~\ref{sec:experiments}. Finally, in Section~\ref{sec:discussion} we provide a discussion and outlook on future work. PRoofs are presented in the appendix.

%%%%%%%%%%%%%%%%%%%%%%%%%%%%%%%%%%%%%%%%%%%%%%%%%%%%%%%%%%%%%%%%%%%%%%%%%
%
%
%
%%%%%%%%%%%%%%%%%%%%%%%%%%%%%%%%%%%%%%%%%%%%%%%%%%%%%%%%%%%%%%%%%%%%%%%%%
\section{Model} \label{sec:model}
An overview of our model, the details of which are given in this section, is given in Figure~\ref{fig:overview}. There is a publicly accessible dataset $\overrightarrow{(s,u)}$ of size $n$, in which each entry $(s_i,u_i)$ is drawn independently from a probability 	distribution $P_{SU}^*$ on a set $\mathcal{S} \times \mathcal{U}$, where $\mathcal{S}$ and $\mathcal{U} \subset \mathbb{R}$ are finite alphabets. New data items $(S,U)$ are also drawn from $P_{SU}^*$. The user's aim is to create a release protocol $P_{Y|SU}$ such that $Y \in \mathcal{U}$ is as close as possible to $U$, while not leaking too much information about $S$. More precisely, the goal is to minimize the distortion, measured as $\mathbb{E}[d(U,Y)]$, where $d$ is any distance function on $\mathcal{U}$.

The distribution $P_{SU}^*$ is not known exactly. The uncertainty set $\mathcal{F}\subset\mathcal{P}_{\mathcal{S}\times\mathcal{U}}$, where $\mathcal{P}_{\mathcal{S}\times\mathcal{U}}$ denotes the probability simplex over $\mathcal{S}\times\mathcal{U}$, captures the user's uncertainty about $P_{SU}^*$. It is constructed from the dataset $\overrightarrow{(s,u)}$. More specifically, we let $\mathcal{F}_B$ be the $(1-\alpha)$-confidence set for $P_{SU}$ in a $\chi^2$-test, i.e.,
\begin{equation} \label{eq:chi2}
\mathcal{F}_B = \left\{P_{SU}\in\mathcal{P}_{\mathcal{S}\times\mathcal{U}}\middle| \sum_{\substack{s \in \mathcal{S},\\u \in \mathcal{U}}} \frac{(\hat{P}_{s,u}-P_{s,u})^2}{P_{s,u}} \leq B \right\},
\end{equation}
where   
\begin{equation}
    B = \frac{F^{-1}_{|\mathcal{S}\times\mathcal{U}|-1}(1-\alpha)}{n},
\end{equation}
$\hat{P}_{SU}$ is the empirical probability distribution of $(S,U)$ and $F_d$ is the CDF of the $\chi^2$-distribution with $d$ degrees of freedom. For the remainder of this paper we fix $B$ and write $\mathcal{F} := \mathcal{F}_B$.

The user creates $P_{Y|SU}$ in such a way that a Local Differential Privacy-like privacy standard is guaranteed when $P_{SU}\in\mathcal{F}$. We will denote this as robust local differential privacy (RLDP) \cite{lopuhaa2021robust}; it is defined as follows.
\begin{definition}
Let $\varepsilon \geq 0$ and $\mathcal{F}\subset\mathcal{P}_{\mathcal{S}\times\mathcal{U}}$. We say that $P_{Y|SU}$ satisfies $(\varepsilon, \mathcal{F})$-RLDP if for all $s_1,s_2 \in \mathcal{S}$, all $y \in \mathcal{Y}$, and all $P_{SU} \in \mathcal{F}$ we have
\begin{equation} \label{eq:rpp}
\mathbb{P}(Y = y|S = s_1) \leq \textrm{\emph{e}}^{\varepsilon}\mathbb{P}(Y = y|S = s_2).
\end{equation}
\end{definition}
    This is less strict than regular LDP \cite{kasiviswanathan2011can}, which requires that not just $S$ but also $U$ be protected. Unlike regular LDP it depends on $P_{SU}$, which is why we demand robustness. In terms of coefficients we can write \eqref{eq:rpp} as
\begin{align}
&\forall y,s_1,s_2 \forall P_{SU} \in \mathcal{F}\colon L_{y,s_1,s_2}(P_{SU},P_{Y|SU}) \leq 0,\\
&L_{y,s_1,s_2}(P_{SU},P_{Y|SU}) \notag\\
&= \sum_{u \in \mathcal{U}}(P_{u|s_1}P_{y|s_1,u}-\textrm{e}^{\varepsilon}P_{u|s_2}P_{y|s_2,u}).
\end{align}

We distinguish four possible optimization problems depending on whether or not we require robustness in i) utility and ii) privacy. This gives:
\begin{align}
\text{\pWW:}\ &\text{minimize}\ \mathbb{E}_{(S,U)\sim \hat P_{SU}}[d(U,Y)] \label{eq:WW} \\
&\text{subject to} \notag \\
  & L_{y,s_1,s_2}(\hat{P}_{SU},P_{Y|SU}) \leq 0, \forall y,s_1,s_2, \notag \displaybreak[0] \\[2mm]
\text{\pWR:}\ &\text{minimize}\ \mathbb{E}_{(S,U)\sim \hat P_{SU}}[d(U,Y)] \label{eq:WR} \\
&\text{subject to} \notag \\
& L_{y,s_1,s_2}(P_{SU},P_{Y|SU}) \leq 0, \notag\\
  &\qquad \forall P_{SU}\in\mathcal{F}, \forall y,s_1,s_2, \notag \displaybreak[0] \\[2mm]
\text{\pRW:}\ &\text{minimize}\ D \label{eq:RW} \\
&\text{subject to} \notag \\
& \mathbb{E}_{(S,U)\sim P_{SU}}[d(U,Y)] \leq D,\quad \forall P_{SU}\in\mathcal{F}, \notag  \\
  & L_{y,s_1,s_2}(\hat{P}_{SU},P_{Y|SU}) \leq 0, \forall y,s_1,s_2, \notag \displaybreak[0] \\[2mm]
\text{\pRR:}\ &\text{minimize}\ D \label{eq:RR} \\
&\text{subject to} \notag \\
& \mathbb{E}_{(S,U)\sim P_{SU}}[d(U,Y)] \leq D,\quad \forall P_{SU}\in\mathcal{F}, \notag  \\
& L_{y,s_1,s_2}(P_{SU},P_{Y|SU}) \leq 0, \forall P_{SU}\in\mathcal{F}, \forall y,s_1,s_2. \notag
\end{align}
% TODO: aligning quantifiers flushed right would be nice
In all these problems the optimization variable is $P_{Y|SU}$. Note that {\pWW} corresponds to the `naive' approach of assuming $P^*_{SU} = \hat{P}_{SU}$ and doing nonrobust optimization.

We note that one can make different choices for the utility \cite{lopuhaa2021privacy}, privacy \cite{makhdoumi2014information,lopuhaa2020privacy}, and uncertainty set \cite{bertsimas2018data}. As we will see below, robust optimization is a general framework that works for many choices, but one needs problem-specific analytic results (Theorems \ref{th:F_support} \& \ref{th:proj_support} and Lemmas \ref{lem:EDstar} \& \ref{lem:tildeLstar} below) in order to reformulate \eqref{eq:WW}--\eqref{eq:RR} as convex optimization problems, which can be fed to a solver.

%%%%%%%%%%%%%%%%%%%%%%%%%%%%%%%%%%%%%%%%%%%%%%%%%%%%%%%%%%%%%%%%%%%%%%%%%
%
%
%
%%%%%%%%%%%%%%%%%%%%%%%%%%%%%%%%%%%%%%%%%%%%%%%%%%%%%%%%%%%%%%%%%%%%%%%%%
\section{Robust optimization} \label{sec:robustopt}
Problems~\eqref{eq:WR}--\eqref{eq:RR} are robust optimization problems~\cite{ben2009robust,ben2015deriving}. In this section we provide some background on robust optimization. 

Let $f: \RR^m\times\RR^n\to\barR$ be defined on the extended real line $\barR=\RR\cup\{-\infty,\infty\}$ and for a given $x\in\RR^n$ define $\dom(f(\cdot\ ,x))=\{a |\ f(a,x)>-\infty \}$. Consider an optimization problem with constraint
\begin{equation} \label{eq:rc}
f(a,x) \leq 0, \forall a\in\mathcal{A},
\end{equation}
where $x\in\RR^n$ is the optimization variable, $a\in\RR^m$ is an uncertain parameter and $\mathcal{A}$ is the uncertainty set for which we need the constraint to hold. The robustness constraints in~\eqref{eq:WR}--\eqref{eq:RR} are of the form~\eqref{eq:rc} with $P_{SU}$ and $P_{Y|SU}$ in the roles of $a$ and $x$, respectively. 

The main tool that is used in this paper is based on Fenchel duality. In order to present this, let
\begin{equation}
f_*(v, x) = \inf_{a\in\mathbb{R}^m} v^Ta - f(a, x)
\end{equation}
and
\begin{equation}
\delta^*(v\given\mathcal{A}) = \sup_{a\in\mathcal{A}} v^T a,
\end{equation}
which are known as the partial concave conjugate of $f$ and the support set of $\mathcal{A}$, respectively. Also, let $\ri(\cdot)$ denote the relative interior of a set. We will extensively use the following result.
\begin{theorem}[\cite{ben2015deriving}] \label{th:rc_dual}
Let $x\in\RR^n$. If $f(a,x)$ is closed concave in $a$ and $\ri(\mathcal{A})\cap\ri(\dom(f(\cdot\ ,x)))\neq\emptyset$, then
$x$ satisfies~\eqref{eq:rc} if and only if 
\begin{equation} \label{eq:rc_dual}
\exists v\in\RR^m \text{ s.t. } \delta^*(v\given\mathcal{A})\leq f_*(v, x).
\end{equation}
\end{theorem}
Since $\delta^*(v\given\mathcal{A})$ and $f_*(v, x)$ are convex and concave functions, respectively, Theorem~\ref{th:rc_dual} provides a means to efficiently solve robust optimization problems once these functions are found in closed form. Another result we use is that Theorem~\ref{th:rc_dual} can be applied independently to each uncertain constraint~\cite{ben2009robust}, and so we can handle robust utility and robust privacy separately.

% The main challenge in applying Theorem~\ref{th:rc_dual} for our problem is that the privacy constraints in~\eqref{eq:WR} and~\eqref{eq:RR} involve $P_{u|s}$, which is a non-convex/concave function of $P_{SU}$. The solution, detailed in the next section, is to work with a projection of $\mathcal{F}$ on these conditional distributions. 

%%%%%%%%%%%%%%%%%%%%%%%%%%%%%%%%%%%%%%%%%%%%%%%%%%%%%%%%%%%%%%%%%%%%%%%%%
%
%
%
%%%%%%%%%%%%%%%%%%%%%%%%%%%%%%%%%%%%%%%%%%%%%%%%%%%%%%%%%%%%%%%%%%%%%%%%%
\section{Closed-form expressions for robustness} \label{sec:results}

While Theorem \ref{th:rc_dual} is a powerful tool to deal with the universal quantifiers in the robust constraints, the downside is that one has to find closed forms of the functions $\delta^*(\bullet|\mathcal{A})$ and $f_*(\bullet,x)$. In this section, we show how to do this for the robust utility and privacy constraints of \eqref{eq:WW}--\eqref{eq:RR}. As will become apparent from the discussion below, the solutions we find are specific to our choice of privacy, utility, and uncertainty set.

\subsection{Robust utility}

The robust utility constraint in {\pRW} and {\pRR} is of the form
\begin{equation} \label{eq:robuti}
\mathbb{E}_{(S,U)\sim P_{SU}}[d(U,Y)] \leq D,\quad \forall P_{SU}\in\mathcal{F}.
\end{equation}
We can write this as
\begin{align}
E_D(P_{SU},P_{Y|SU}) &\leq 0, \quad \forall P_{SU} \in \mathcal{F},\\
E_D(P_{SU},P_{Y|SU}) &= \mathbb{E}[d(U,Y)]-D\\
&= \sum_{y,s,u} P_{s,u} P_{y|s,u} d(u,y)-D.
\end{align}
Using Theorem \ref{th:rc_dual} we can write this as
\begin{equation}
\exists v \in \mathbb{R}^{|\mathcal{S}\times\mathcal{U}|}\colon \delta^*(v|\mathcal{F}) \leq E_{D*}(v,P_{Y|SU}).
\end{equation}

Thus we need to find expressions for $\delta^*(v|\mathcal{F})$ and $E_{D*}(v,P_{Y|SU})$. Since $E_D$ is linear in $P_{SU}$, one easily derives the following \cite{ben2009robust}:

\begin{lemma} \label{lem:EDstar}
Let $E_D(P_{SU},P_{Y|SU})=\EE[d(U,Y)]-D$, with uncertain parameter $P_{SU}$ and optimization variable $P_{Y|SU}$. Then
\begin{align}
&E_{D*}(v, P_{Y|SU}) \notag \\
&=
\begin{cases}
D, &\text{if } \forall s,u\colon v_{s,u}=\sum_{y} P_{y|s,u} d(y,u), \\
-\infty,\quad &\text{otherwise.}
\end{cases}
\end{align}
\end{lemma}

The expression for $\delta^*(v|\mathcal{F})$ is a little bit more work, but can be found using established results in convex analysis \cite{ben2015deriving}. A related result is found in \cite[Thm.~5]{bertsimas2018data}.

\begin{theorem} \label{th:F_support}
\begin{multline}
\delta^*( v |  \mathcal{F} ) = \min_{w\geq v ,c\geq 0}\Bigg\{ -2\sqrt{c}\sum_{s,u}\hat P_{s,u}\sqrt{w_{s,u}-v_{s,u}} \\
+ \max_{s,u} w_{s,u}  + c(B+1)\Bigg\}.
\end{multline}
\end{theorem}
Combining these results, we can write the robust utility constraint \eqref{eq:robuti} as
\begin{equation} \label{eq:robuti2}
\delta^*\left(\left(\sum_y P_{y|s,u}d(y,u)\right)_{s \in \mathcal{S},u \in \mathcal{U}} \middle| \mathcal{F}\right) \leq D,
\end{equation}
where $\delta^*(\bullet|\mathcal{F})$ is as in Theorem \ref{th:F_support}.

\subsection{Robust privacy}

Fix $y,s_1,s_2$. The robust privacy constraint in {\pWR} and {\pRR} is of the form
\begin{equation} \label{eq:robpriv}
L_{y,s_1,s_2}(P_{SU},P_{Y|SU}) \leq 0, \forall P_{SU} \in \mathcal{F}.    
\end{equation}
One would like to apply Theorem \ref{th:rc_dual}; however, $L_{y,s_1.s_2}$ is not convex in $P_{SU}$. Therefore we rewrite it as follows. Let $\mathcal{F}_{U|s_1,s_2}$ be the projection of $\mathcal{F}$ onto $\mathcal{P}_{\mathcal{U}}\times\mathcal{P}_{\mathcal{U}}$ by the map $P\mapsto(P_{U|s_1},P_{U|s_2})$. We can write \eqref{eq:robpriv} as
\begin{align}
&\forall (P_{U|s_1},P_{U|s_2}) \in \mathcal{F}_{U|s_1,s_2}\colon \notag \\
&\tilde{L}_{y,s_1,s_2}(P_{U|s_1},P_{U|s_2},P_{Y|SU}) \leq 0,
\end{align}
where
\begin{align}
&\tilde{L}_{y,s_1,s_2}(P_{U|s_1},P_{U|s_2},P_{Y|SU}) \notag \\
&= \sum_{u \in \mathcal{U}}(P_{u|s_1}P_{y|s_1,u}-\textrm{e}^{\varepsilon}P_{u|s_2}P_{y|s_2,u}).
\end{align}
Again we use Theorem \ref{th:rc_dual} to write this as
\begin{equation}
\exists v_1,v_2 \in \mathbb{R}^{|\mathcal{U}|}\colon \delta^*(v_1,v_2|\mathcal{F}_{U|s_1,s_2}) \leq \tilde{L}_{y,s_1,s_2*}(v_1,v_2,P_{Y|SU}).
\end{equation}
Thus we need to find expressions for these functions. As $\tilde{L}_{y,s_1,s_2}(P_{U|s_1},P_{U|s_2},P_{Y|SU})$ is linear in its first two arguments, we find the following:

\begin{lemma} \label{lem:tildeLstar}
One has
\begin{align}
&\tilde{L}_{y,s_1,s_2*}(v_1,v_2, P_{Y|SU}) \notag \\
&=
\begin{cases}
0, &\text{if } \forall u\colon  v_{1,u} = P_{y|s_1,u},\\
	&\quad \text{and } v_{2,u} = -\textrm{\emph{e}}^\epsilon P_{y|s_2,u}, \\
-\infty,\ &\text{otherwise.}
\end{cases}
\end{align}
\end{lemma}

The expression for $\delta^*(v_1,v_2|\mathcal{F}_{U|s_1,s_2})$ takes considerably more effort, in part because we first have to find a closed expression for $\mathcal{F}_{U|s_1,s_2}$. This is given in the following result:

\begin{theorem} \label{th:proj_form}
\begin{multline} \label{eq:fuss}
\mathcal{F}_{U|s_1,s_2} = \Bigg\{ \left(R_{U|s_1},R_{U|s_2}\right)\in\mathcal{P}_{\mathcal{U}}^2\ \Bigg| \\
\sum_{i=1}^2\sqrt{\sum_u \frac{\hat P_{s_i,u}^2}{R_{u|s_i}}} 
\leq
\sqrt{B+1} - 1 + \sum_{i=1}^2\hat P_{s_i} 
 \Bigg\}.
\end{multline}
\end{theorem}

Similar to, but more in a more complicated way than, Theorem \ref{th:F_support}, one can prove:

\begin{theorem} \label{th:proj_support}
The support function of $\mathcal{F}_{U|s_1,s_2}$ is given by
% DISCUSSION: maybe explicitly introduce domain of v_1, v_2, ...
\begin{multline}
\delta^*(v_1,v_2|\mathcal{F}_{U|s_1,s_2}) =
\min_{c\geq 0,w_1\geq v_1,w_2\geq v_2}
\Bigg\{ \\
-(2^{-2/3}+2^{1/3})c^{2/3}\sum_{i=1}^2\left(\sum_{u}\hat P_{s_i,u}\sqrt{w_{i}(u)-v_{i}(u)}\right)^{2/3}
 \\
+ \sum_{i=1}^2\max_u w_{i}(u) 
+ c\bigg(\sqrt{B+1} - 1 + \sum_{i=1}^2\hat P_{s_i}
\bigg)
\Bigg\}.
\end{multline}
\end{theorem}
Thus we can write \eqref{eq:robpriv} as
\begin{equation} \label{eq:robpriv2}
\delta^*\left((P_{y|s_1,u})_{u \in \mathcal{U}},(P_{y|s_2,u})_{u \in \mathcal{U}}\middle|\mathcal{F}_{\mathcal{U}|s_1,s_2}\right) \leq 0,
\end{equation}
where $\delta^*(\bullet|\mathcal{F}_{\mathcal{U}|s_1,s_2})$ is as in Theorem \ref{th:proj_support}. Replacing \eqref{eq:robuti} by \eqref{eq:robuti2}, and \eqref{eq:robpriv} by \eqref{eq:robpriv2} allows us to provide convex formulations of~\eqref{eq:WW}--\eqref{eq:RR}. In the next section we provide some insights that are obtained through numerical experiments.

\begin{figure*}
\centering
\begin{subfigure}{0.45\linewidth}
    \centering
\begin{tikzpicture}
\begin{axis}[
  xlabel=$\varepsilon^*$,ylabel=$D^*$,
  font=\scriptsize,
  legend style={
        cells={anchor=west},
        legend pos=north east,
       font=\scriptsize,
    }
]

\addplot[
  color=orange, only marks,
  mark=square*,mark size=.4mm,mark options={solid}
  ]
table[
  header=false,x index=0,y index=1,
  ]
{./data_fig1.csv};
\addlegendentry{\pWW};

\addplot[
  color=lightgray, only marks,
  mark=diamond*,mark size=.6mm,mark options={solid}
  ]
table[
  header=false,x index=2,y index=3,
  ]
{./data_fig1.csv};
\addlegendentry{\pWR};

\addplot[
  color=cyan, only marks,
  mark=triangle*,mark size=.5mm,mark options={solid}
  ]
table[
  header=false,x index=4,y index=5,
  ]
{./data_fig1.csv};
\addlegendentry{\pRW};

\addplot[
  color=teal, only marks,
  mark=*,mark size=.5mm,mark options={solid}
  ]
table[
  header=false,x index=6, y index=7,
  ]
{./data_fig1.csv};
\addlegendentry{\pRR};

\end{axis}
\end{tikzpicture}
\caption{Impact of robustness on utility ($K=30$, $n=75$).  \label{fig:scatter1}}
\end{subfigure} \ \ 
\begin{subfigure}{0.45\linewidth}
    \centering
\begin{tikzpicture}
\begin{axis}[
  xlabel=$\varepsilon^*$,ylabel=$D^*$,
  font=\scriptsize,
  legend style={
        cells={anchor=west},
        legend pos=north east,
       font=\scriptsize,
    }
]

\addplot[
  color=orange, only marks,
  mark=square*,mark size=.4mm,mark options={solid}
  ]
table[
  header=false,x index=0,y index=1,
  ]
{./data_fig2.csv};
\addlegendentry{\pWW};

\addplot[
  color=lightgray, only marks,
  mark=diamond*,mark size=.6mm,mark options={solid}
  ]
table[
  header=false,x index=2,y index=3,
  ]
{./data_fig2.csv};
\addlegendentry{\pWR};

\addplot[
  color=cyan, only marks,
  mark=triangle,mark size=.7mm,mark options={solid}
  ]
table[
  header=false,x index=4,y index=5,
  ]
{./data_fig2.csv};
\addlegendentry{\pRW};

\addplot[
  color=teal, only marks,
  mark=o,mark size=.6mm,mark options={solid}
  ]
table[
  header=false,x index=6, y index=7,
  ]
{./data_fig2.csv};
\addlegendentry{\pRR};

\end{axis}
\end{tikzpicture}

\caption{Impact of robustness on utility ($K=30$, $n=15,000$). \label{fig:scatter2}}
\end{subfigure}

\begin{subfigure}{0.45\linewidth}
    \centering
\begin{tikzpicture}
\begin{semilogxaxis}[
  xlabel=$N$,ylabel=$\varepsilon^*$,
  font=\scriptsize,
  legend style={
        cells={anchor=west},
        legend pos=north east,
       font=\scriptsize,
    }
]

%%%%%%%%%%%%%%%%%
%
% WW
%
%%%%%%%%%%%%%%%%%
\addplot[
  color=orange,
  line width=.5mm
  ]
table[
  header=false,x index=0, y index=1,
  ]
{./data_fig3.csv};

\addplot[
  color=orange,
  name path = WWtop
  ]
table[
  header=false,x index=0, y index=9,
  ]
{./data_fig3.csv};

\addplot[
  color=orange,
  name path = WWbottom
  ]
table[
  header=false,x index=0, y index=17,
  ]
{./data_fig3.csv};

\addplot[orange!20] fill between[of=WWtop and WWbottom];

%%%%%%%%%%%%%%%%%
%
% WR
%
%%%%%%%%%%%%%%%%%
\addplot[
  color=lightgray,
  line width=.5mm
  ]
table[
  header=false,x index=0, y index=2,
  ]
{./data_fig3.csv};

\addplot[
  color=lightgray,
  name path = WRtop
  ]
table[
  header=false,x index=0, y index=10,
  ]
{./data_fig3.csv};

\addplot[
  color=lightgray,
  name path = WRbottom
  ]
table[
  header=false,x index=0, y index=18,
  ]
{./data_fig3.csv};

\addplot[lightgray!20] fill between[of=WRtop and WRbottom];

%%%%%%%%%%%%%%%%%
%
% RW
%
%%%%%%%%%%%%%%%%%
\addplot[
  color=cyan,
  line width=.5mm
  ]
table[
  header=false,x index=0, y index=3,
  ]
{./data_fig3.csv};

\addplot[
  color=cyan,
  name path = RWtop
  ]
table[
  header=false,x index=0, y index=11,
  ]
{./data_fig3.csv};

\addplot[
  color=cyan,
  name path = RWbottom
  ]
table[
  header=false,x index=0, y index=19,
  ]
{./data_fig3.csv};

\addplot[cyan!20] fill between[of=RWtop and RWbottom];

%%%%%%%%%%%%%%%%%
%
% RR
%
%%%%%%%%%%%%%%%%%
\addplot[
  color=teal,
  line width=.5mm
  ]
table[
  header=false,x index=0, y index=4,
  ]
{./data_fig3.csv};

\addplot[
  color=teal,
  name path = RRtop
  ]
table[
  header=false,x index=0, y index=12,
  ]
{./data_fig3.csv};

\addplot[
  color=teal,
  name path = RRbottom
  ]
table[
  header=false,x index=0, y index=20,
  ]
{./data_fig3.csv};

\addplot[teal!20] fill between[of=RRtop and RRbottom];

\legend{\pWW,,,,\pWR,,,,\pRW,,,,\pRR,,,};

\end{semilogxaxis}
\end{tikzpicture}

\caption{This figure depicts $\varepsilon^*$ as a function of $n=N\cdot|\mathcal{S}| |\mathcal{U}|$. \label{fig:vary_in_N_epsilon_star}}
\end{subfigure} \ \ 
\begin{subfigure}{0.45\linewidth}
    \centering
\begin{tikzpicture}
\begin{semilogxaxis}[
%   width=8cm,
  %xmode = log
  xlabel=$N$,ylabel=$D^*$,
  font=\scriptsize,
  legend style={
        cells={anchor=west},
        legend pos=north east,
       font=\scriptsize,
    }
]

%%%%%%%%%%%%%%%%%
%
% WW
%
%%%%%%%%%%%%%%%%%
\addplot[
  color=orange,
  line width = 0.5mm
  ]
table[
  header=false,x index=0, y index=5,
  ]
{./data_fig3.csv};

\addplot[
  color=orange,
  name path = WWtop
  ]
table[
  header=false,x index=0, y index=13,
  ]
{./data_fig3.csv};

\addplot[
  color=orange,
  name path = WWbottom
  ]
table[
  header=false,x index=0, y index=21,
  ]
{./data_fig3.csv};

\addplot[orange!20] fill between[of=WWtop and WWbottom];

%%%%%%%%%%%%%%%%%
%
% WR
%
%%%%%%%%%%%%%%%%%
\addplot[
  color=lightgray,
  line width = 0.5mm
  ]
table[
  header=false,x index=0, y index=6,
  ]
{./data_fig3.csv};

\addplot[
  color=lightgray,
  name path = WRtop
  ]
table[
  header=false,x index=0, y index=14,
  ]
{./data_fig3.csv};

\addplot[
  color=lightgray,
  name path = WRbottom
  ]
table[
  header=false,x index=0, y index=22,
  ]
{./data_fig3.csv};

\addplot[lightgray!20] fill between[of=WRtop and WRbottom];

%%%%%%%%%%%%%%%%%
%
% RW
%
%%%%%%%%%%%%%%%%%
\addplot[
  color=cyan,
  line width = 0.5mm
  ]
table[
  header=false,x index=0, y index=7,
  ]
{./data_fig3.csv};

\addplot[
  color=cyan,
  name path = RWtop
  ]
table[
  header=false,x index=0, y index=15,
  ]
{./data_fig3.csv};

\addplot[
  color=cyan,
  name path = RWbottom
  ]
table[
  header=false,x index=0, y index=23,
  ]
{./data_fig3.csv};

\addplot[cyan!20] fill between[of=RWtop and RWbottom];

%%%%%%%%%%%%%%%%%
%
% RR
%
%%%%%%%%%%%%%%%%%
\addplot[
  color=teal,
  line width = 0.5mm
  ]
table[
  header=false,x index=0, y index=8,
  ]
{./data_fig3.csv};

\addplot[
  color=teal,
  name path = RRtop
  ]
table[
  header=false,x index=0, y index=16,
  ]
{./data_fig3.csv};

\addplot[
  color=teal,
  name path = RRbottom
  ]
table[
  header=false,x index=0, y index=24,
  ]
{./data_fig3.csv};

\addplot[teal!20] fill between[of=RRtop and RRbottom];

\legend{\pWW,,,,\pWR,,,,\pRW,,,,\pRR,,,};

\end{semilogxaxis}
\end{tikzpicture}

\caption{This figure depicts $D^*$ as a function of $n=N\cdot|\mathcal{S}| |\mathcal{U}|$. \label{fig:vary_in_N_D}}
\end{subfigure}
\caption{Experimental results on $K$ instances for optimization under {\sffamily N}onrobust/{\sffamily R}obust {\sffamily U}tility and {\sffamily N}onrobust/{\sffamily R}obust {\sffamily P}rivacy. ($\alpha=0.05$, $\varepsilon=0.5$, $|\mathcal{S}| = 3$, $|\mathcal{U}| = 5$) \label{fig:all}}
\end{figure*}
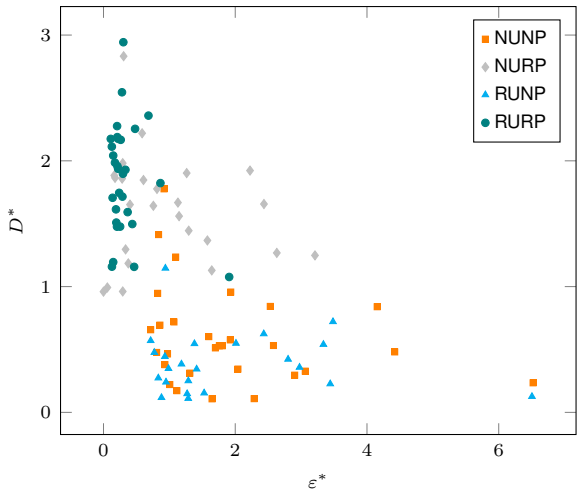
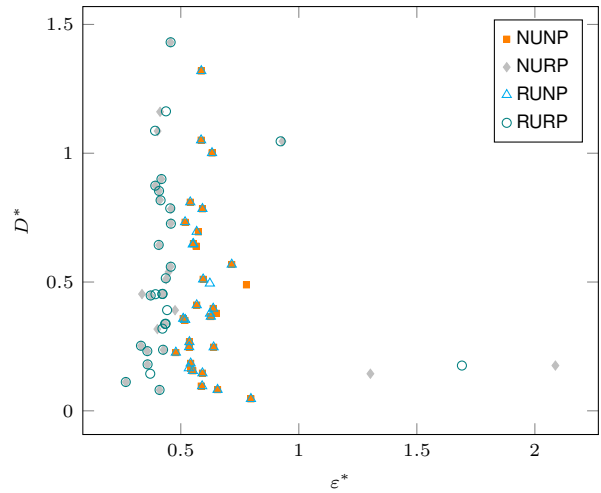
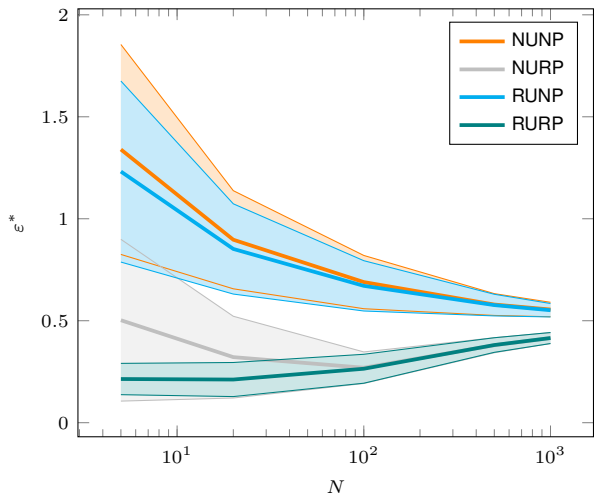
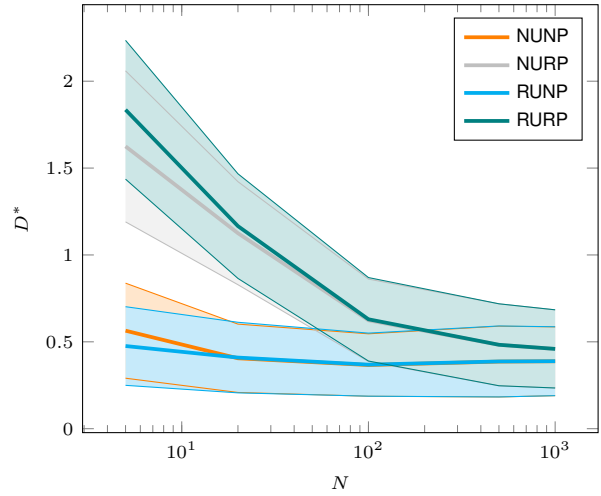

\section{Numerical experiments} \label{sec:experiments}
%We will distinguish in the amount of robustness that is desired for utility and privacy by introducing $\alpha_d$ and $\alpha_p$, respectively. As such we consider different uncertainty sets (different values of $B$ in~\eqref{eq:chi2}) for the utility and privacy constraints in~\eqref{eq:WW}--\eqref{eq:RR}.

The performance measures that we consider are 
\begin{equation}
    D^* = \EE_{(S,U)\sim P^*_{SU}}[d(U,Y)]
\end{equation}
and
\begin{equation}
    \varepsilon^* = \log\max_{y,s_1,s_2}\frac{P_{y|s_1}}{P_{y|s_2}},
\end{equation}
interpreting $0/0$ as $1$ and
with $\varepsilon^*=\infty$ if there exist $y$, $s_1$ and $s_2$ for which $P_{y|s_2}=0$ and $P_{y|s_1}>0$. These measures give the performance of the $P_{Y|SU}$ that is obtained from optimization under the actual distribution $P_{SU}^*$.

In all experiments we use $\mathcal{S}=\{0,1,2\}$, $\mathcal{U}=\{0,1,2,3,4\}$, $d(x,y) = (x-y)^2$, $\alpha = 0.05$ and $\varepsilon = 0.5$. We draw $P_{SU}^*$ according to the Jeffreys prior on $\mathcal{P}_{\mathcal{S}\times\mathcal{U}}$, i.e., the symmetric Dirichlet distribution with parameter $\tfrac{1}{2}$ \cite{robert2009harold}.

In our first experiment we draw $K=30$ instances of $P_{SU}^*$. For each instance, we draw $n$ samples $(s,u)$ from which we estimate $\hat P_{SU}$. For each of the $K=30$ instances we report $\varepsilon^*$ and $D^*$ in a scatter plot. We do so for each combination of nonrobust/robust utility and privacy (nomenclature in \eqref{eq:WW}--\eqref{eq:RR}). In Figures~\ref{fig:scatter1} and~\ref{fig:scatter2} the results are depicted for $n=5|\mathcal{S}| |\mathcal{U}|=75$ and for $n=10^3|\mathcal{S}| |\mathcal{U}|=15,000$, respectively. Instances that have $\epsilon^*=\infty$ are omitted from the figures.

We observe that without robustness in privacy, i.e., {\pWW} and {\pRW}, almost all instances have significantly lower privacy performance than what they are designed for, i.e., $\varepsilon^*\gg\varepsilon$. Also, by introducing robustness in privacy as a hard constraint, we significantly suffer in terms of utility. Note though that even with robust privacy we do not always get $\varepsilon^* \leq \varepsilon$; this is because $P^* \in \mathcal{F}$ only with probability $1-\alpha$. In Figure \ref{fig:scatter1} {\pWR} has more than the expected number of outliers, which can be explained from the fact that the $\chi^2$ test only asymptotically gives a confidence interval as $n \rightarrow \infty$. Another interesting observation (which will be confirmed by the next experiments) is that once we have imposed robust privacy, imposing robust utility does not make much difference, i.e., {\pWR} and {\pRR} are very similar. This shows that the utility cost of demanding robust privacy is considerably greater than the utility benefit of demanding robust utility. Note, that by comparing to Figure~\ref{fig:scatter2}, some of the large $\varepsilon^*$ values for {\pWR} in Figure~\ref{fig:scatter1} seem to be small sample artifacts. A final observation is that in both figures there are outliers, especially in $\varepsilon^*$. 

Our second experiment illustrates the influence of $n$. We give the mean $D^*$ and $\varepsilon^*$ over $K=10^3$ instances as a function of $n$ in Figures~\ref{fig:vary_in_N_epsilon_star} and~\ref{fig:vary_in_N_D}, respectively. In addition to the mean we report the standard deviation. Thick lines depict the mean over $K=10^3$ samples, the width of the bands correspond to $\pm 1$ standard deviation. Both are computed by first removing outliers based on the $1.5\text{IQR}$ rule.

We observe that for small $n$, due to the large uncertainty on $P_{SU}$, {\pRR} is very conservative with $\varepsilon^*\ll\epsilon$. The figures also confirm that imposing robustness in utility has relatively little impact.

%Our third and final experiment demonstrates the influence of $\alpha_p$. The results, with $\epsilon=0.5$, $\alpha_d=0.5$, $N=$ and $K=500$, are depicted in Figures~\ref{fig:vary_in_alpha_p_epsilon_star} and~\ref{fig:vary_in_alpha_p_D}. We observe that as $\alpha_p$ increases, i.e. as the size of the uncertainty set decreases, the impact of imposing robustness decreases.
% FIGURE
%\input{fig_epsilon_star_in_alpha_p}
% FIGURE
%\input{fig_D_in_alpha_p}

Our implementations are in CVX~\cite{cvx} and are solved using Mosek~\cite{mosek}.

%%%%%%%%%%%%%%%%%%%%%%%%%%%%%%%%%%%%%%%%%%%%%%%%%%%%%%
%
%
%
%%%%%%%%%%%%%%%%%%%%%%%%%%%%%%%%%%%%%%%%%%%%%%%%%%%%%%
\section{Discussion} \label{sec:discussion}
We have given convex formulations of optimization problems for finding robust data release protocols $P_{Y|SU}$. In addition we also studied the non-robust form. Numerical experiments revealed that the non-robust forms achieve privacy levels that are much worse than anticipated. In particular, the naive approach of assuming $P^*_{SU} = \hat{P}_{SU}$ leads to undesirable privacy leakage.

Our convex formulations and corresponding implementations have a number of variables and constraints that grows exponentially in $|\mathcal{S}|$ and $|\mathcal{U}|$. It would be of great interest to develop bounding methods that have reduced complexity, but that still provide strong guarantees. 

The current model imposes that the output alphabet of $P_{Y|SU}$ is equal to the input alphabet $\mathcal{U}$. It would be of interest to leverage this assumption.

%%%%%%%%%%%%%%%%%%%%%%%%%%%%%%%%%%%%%%%%%%%%%%%%%%%%%%
%
%
%
%%%%%%%%%%%%%%%%%%%%%%%%%%%%%%%%%%%%%%%%%%%%%%%%%%%%%%

%%%%%%%%%%%%%%%%%%%%%%%%%%%%%%%%%%%%%%%%%%%%%%%%%%%%%%
%
%
%
%%%%%%%%%%%%%%%%%%%%%%%%%%%%%%%%%%%%%%%%%%%%%%%%%%%%%%
\bibliographystyle{IEEEtran}
\bibliography{IEEEabrv,robustsldp}

\appendices

%%%%%%%%%%%%%%%%%%%%%%%%%%%%%%%%%%%%%%%%%%%%%%%%%%%%%%
%
%
%
%%%%%%%%%%%%%%%%%%%%%%%%%%%%%%%%%%%%%%%%%%%%%%%%%%%%%%
\section{Proof of Theorem~\ref{th:proj_form}} \label{app:form}
\begin{IEEEproof}
Denote the RHS of \eqref{eq:fuss} by $\overline{\mathcal{F}}_{U|s_1,s_2}$. Let $ \left(R_{U|s_1},R_{U|s_2}\right)\in\overline{\mathcal{F}}_{U|s_1,s_2}$ and define $P_{SU}\in\mathbb{R}^\mathcal{S}\times\mathbb{R}^\mathcal{U}$ as
\begin{equation}
P_{s,u} =
\begin{cases}
\frac{\kappa_1}{\kappa_1+\kappa_2+\kappa_3}R_{u|s_1}, \quad &\text{ if } s = s_1, \\
\frac{\kappa_2}{\kappa_1+\kappa_2+\kappa_3}R_{u|s_2}, \quad &\text{ if } s = s_2, \\
\frac{1}{\kappa_1+\kappa_2+\kappa_3}\hat P_{s,u}, \quad &\text{ if } s \neq s_1, s_2, \\
\end{cases}
\end{equation}
with
\begin{equation}
\kappa_1 = \hat P_{s_1}\sqrt{\sum_u \frac{\hat P_{u|s_1}^2}{R_{u|s_1}}}, \quad
\kappa_2 = \hat P_{s_2}\sqrt{\sum_u \frac{\hat P_{u|s_2}^2}{R_{u|s_2}}},
\end{equation}
and $\kappa_3 = 1 - \hat P_{s_1} - \hat P_{s_2}$. Then
\begin{align}
\sum_{s,u} P_{s,u} &= \tfrac{1}{\kappa_1+\kappa_2+\kappa_3}\Bigg(\\
&\kappa_1 \sum_u R_{u|s_1}+\kappa_2 \sum_u R_{u|s_2} + \sum_{s \notin \{s_1,s_2\},u} \hat{P}_{s,u}\Bigg) \notag\\
&= \tfrac{1}{\kappa_1+\kappa_2+\kappa_3}(\kappa_1+\kappa_2+\kappa_3) = 1,
\end{align}
so $P_{SU} \in \mathcal{P}_{\mathcal{S}\times\mathcal{U}}$. Furthermore,
\begin{align}
\sum_{s,u}\frac{P_{s,u}^2}{P_{s,u}} &= (\kappa_1+\kappa_2+\kappa_3)\Bigg( \\ &\sum_u\frac{\hat{P}_{s,u}^2}{\kappa_1R_{u|s_1}}+\frac{\hat{P}_{s,u}^2}{\kappa_2R_{u|s_2}}+\sum_{s \notin \{s_1,s_2\},u} \hat{P}_{s,u}\Bigg) \notag\\
&= (\kappa_1+\kappa_2+\kappa_3)^2.
\end{align}
Since $(R_{U|s_1},R_{U|s_2} \in \overline{\mathcal{F}}_{U|s_1,s_2}$, one has $\kappa_1+\kappa_2+\kappa_3 \leq \sqrt{B+1}$. It follows that 
\begin{align}
\sum_{s,u}\frac{(P_{s,u}-\hat{P}_{s,u})^2}{P_{s,u}} &= \sum_{s,u} (P_{s,u}-2\hat{P}_{s,u}) + \sum_{s,u}\frac{P_{s,u}^2}{P_{s,u}} \\
&= (\kappa_1+\kappa_2+\kappa_3)^2-1 \\
&\leq B,
\end{align}
which shows that $P_{SU}\in\mathcal{F}$. It is clear that $P_{SU}$ projects to $\left(R_{U|s_1},R_{U|s_2}\right)$. This shows that $\overline{\mathcal{F}}_{U|s_1,s_2} \subset \mathcal{F}_{U|s_1,s_2}$.

Next, let $P_{SU}\in\mathcal{F}$. From~\eqref{eq:chi2} it follows that
\begin{align} 
\frac{\hat P_{s_1}^2}{P_{s_1}}\sum_{u}\frac{\hat P_{u|s_1}^2}{P_{u|s_1}} + \frac{\hat P_{s_2}^2}{P_{s_2}}\sum_{u}\frac{\hat P_{u|s_2}^2}{P_{u|s_2}} +&  \notag\\
\frac{\left(1-\hat P_{s_1} - \hat P_{s_2}\right)^2}{1-P_{s_1}-P_{s_2}} &\leq B+1. \label{eq:doubleproj_a}
\end{align}
Application of the Cauchy-Schwartz inequality on the vectors
\begin{equation*}
\left( \frac{\hat P_{s_1}}{\sqrt{P_{s_1}}}\sqrt{\sum_u \frac{\hat P_{u|s_1}^2}{P_{u|s_1}}}, \frac{\hat P_{s_2}}{\sqrt{P_{s_2}}}\sqrt{\sum_u \frac{\hat P_{u|s_2}^2}{P_{u|s_2}}}, \frac{1 - \hat P_{s_1} - \hat P_{s_2}}{\sqrt{1-P_{s_1}-P_{s_2}}} \right)
\end{equation*}
and
\begin{equation*}
\left(\sqrt{P_{s_1}}, \sqrt{P_{s_2}}, \sqrt{1-P_{s_1}-P_{s_2}}\right),
\end{equation*}
gives
\begin{align}
&\left(\hat P_{s_1}\sqrt{\sum_u \frac{\hat P_{u|s_1}^2}{P_{u|s_1}}} + \hat P_{s_2}\sqrt{\sum_u \frac{\hat P_{u|s_2}^2}{P_{u|s_2}}} + 1 - \hat P_{s_1} - \hat P_{s_2} \right)^2 \notag\\
&\leq
\frac{\hat P_{s_1}^2}{P_{s_1}}\sum_{u}\frac{\hat P_{u|s_1}^2}{P_{u|s_1}}+ \frac{\hat P_{s_2}^2}{P_{s_2}}\sum_{u}\frac{\hat P_{u|s_2}^2}{P_{u|s_2}}\\
&\quad + \frac{\left(1-\hat P_{s_1} - \hat P_{s_2}\right)^2}{1-P_{s_1}-P_{s_2}}, \notag
\end{align}
which together with~\eqref{eq:doubleproj_a} yields
\begin{align}
\hat P_{s_1}\sqrt{\sum_u \frac{\hat P_{u|s_1}^2}{P_{u|s_1}}} + \hat P_{s_2}\sqrt{\sum_u \frac{\hat P_{u|s_2}^2}{P_{u|s_2}}}& \notag \\
+ 1 - \hat P_{s_1} - \hat P_{s_2} &\leq \sqrt{B+1},
\end{align}
which shows that $\mathcal{F}_{U|s_1,s_2} \subset \overline{\mathcal{F}}_{U|s_1,s_2}$. This completes the proof.
\end{IEEEproof}

\section{Concave conjugates}

In this section we prove Lemmas \ref{lem:EDstar} and \ref{lem:tildeLstar}. Since both $E_D$ and $\tilde{L}_{y,s_1,s_2}$ are linear in the uncertain variable, they both follow directly from the following result:

\begin{lemma}
Let $f$ be a convex function, and let $x$ be such that $f(a,x)$ is linear in $v$; write $f(v,x) = b_x^Ta + c_x$. Then
\begin{equation}
f_*(v,x) = \begin{cases}
-c_x, & \text{ if } v = b_x,\\
-\infty, & \text{ otherwise.}
\end{cases}
\end{equation}
\end{lemma}

\begin{IEEEproof}
By definition
\begin{equation}
f_*(v,x) = \inf_a (v-b_x)^Ta-c_x.
\end{equation}
If $v-b_x \neq 0$, then the inner product can become arbitrarily negative, and so $f_*(v,x) = -\infty$. If $v = b_x$, however, then the RHS is equal to $-c_x$ no matter the choice of $a$.
\end{IEEEproof}
`
\section{Support functions}

In this appendix we prove Theorems \ref{th:F_support} and \ref{th:proj_support}. We first state a number of standard results of convex analysis, which can be found, for instance, in \cite{ben2015deriving}. For a closed convex function $f\colon \mathbb{R}^n \rightarrow \mathbb{R} \cup \{\pm \infty\}$ and $v \in \mathbb{R}^n$ we write
\begin{equation}
f^*(v) = \sup_{x \in \mathbb{R}^n} v^Tx-f(x).
\end{equation}

\begin{lemma}[\cite{ben2015deriving}] \label{lem:appconst}
For $C \in \mathbb{R}$ one has $(f+C)^*(v) = f^*(v)-C$.
\end{lemma}

\begin{lemma}[\cite{ben2015deriving}] \label{lem:appsum}
Let $h_1,\ldots,h_k \colon \mathbb{R}^n \rightarrow \mathbb{R} \cup \pm{\infty}$ be closed convex functions, and let \begin{equation}
\mathcal{A} = \left\{(x_1,\ldots,x_k) \in \mathbb{R}^{2n} \middle| \sum_j h_j(x_j) \leq 0\right\}.
\end{equation}
Then
\begin{equation}
\delta^*(v_1,\ldots,v_k|\mathcal{A}) = \min_{c \geq 0}\left\{c\sum_j h_j^*\left(\tfrac{v_j}{c}\right)\right\}.
\end{equation}
\end{lemma}

\begin{lemma}[\cite{ben2015deriving}] \label{lem:appcart}
One has $\delta^*(v_1,v_2|\mathcal{A} \times \mathcal{B}) = \delta^*(v_1|\mathcal{A}) + \delta^*(v_2|\mathcal{B})$.
\end{lemma}

\begin{lemma}[\cite{ben2015deriving}] \label{lem:appint}
Let $\mathcal{A},\mathcal{B}$ be closed sets such that $\operatorname{ri}(\mathcal{A}) \cap \operatorname{ri}(\mathcal{B}) \neq \varnothing$. Then
\begin{equation}
\delta^*(v|\mathcal{A} \cap \mathcal{B}) = \min_{t,w}\left\{\delta^*(t|\mathcal{A})+\delta^*(w|\mathcal{B})\middle| t+w = v\right\}.
\end{equation}
\end{lemma}

\begin{IEEEproof}[Proof of Theorem \ref{th:F_support}]
We can write
\begin{align}
\delta^*(v|\mathcal{F}) &= \delta^*(v|\mathcal{P}_{\mathcal{S}\times\mathcal{U}} \cap \mathcal{B}),\\
\mathcal{B} &= \{R \in \mathbb{R}^{|\mathcal{S} \times \mathcal{U}|} | g(R)-B-1 \leq 0\},\\
g(R) &= \sum_{s,u} g_{s,u}(R_{s,u}),\\
g_{s,u}(R_{s,u}) &= \frac{\hat{P}_{s,u}^2}{R_{s,u}}.
\end{align}
Then by Lemma \ref{lem:appint}
\begin{align}
\delta^*(v|\mathcal{F}) = \min_{t,v} \{\delta^*(t|\mathcal{B}) + \delta^*(w|\mathcal{P}_{\mathcal{S} \times \mathcal{U}}) \mid t+w=v\}. \label{eq:appf1}
\end{align}
For the second term on the RHS have 
\begin{align}
\delta^*(w|\mathcal{P}_{\mathcal{S} \times \mathcal{U}}) &= \max_{P_{SU} \in \mathcal{P}_{\mathcal{S}\times\mathcal{U}}} \sum_{s,u} P_{s,u}w_{s,u}
&= \max_{s,u} w_{s,u}. \label{eq:appf2}
\end{align}
Furthermore it follows from Lemmas \ref{lem:appconst} and \ref{lem:appsum} that
\begin{align}
\delta^*(t|\mathcal{B}) &= \min_{c\geq 0}\left\{c\sum_{s,u} g^*_{s,u}\left(\tfrac{t_{s,u}}{c}\right)+c(B+1)\right\},
\end{align}
so it remains to determine
\begin{align}
g^*_{s,u}(z) = \sup_x xz - \frac{\hat{P}_{s,u}^2}{x}.
\end{align}
for $z \in \mathbb{R}$. We find this by taking the derivative w.r.t. $x$, and we have to solve
\begin{equation}
z + \frac{\hat{P}_{s,u}^2}{x^2} = 0
\end{equation}
hence $x = \frac{\hat{P}_{s,u}^2}{\sqrt{-z}}$ (if $z > 0$, then the maximum does not exist and $g^*_{s,u}(z) = \infty$). Substituting this we find
\begin{equation}
g^*_{s,u}(z) = -2\sqrt{-z}\hat{P}_{s,u}.
\end{equation}
Combining this with \eqref{eq:appf1} and \eqref{eq:appf2}, and substituting $t = w-v$, now proves the Theorem.
\end{IEEEproof}

\begin{IEEEproof}[Proof of Theorem \ref{th:proj_form}]
We have
\begin{align}
\delta^*(v_1,v_2|\mathcal{F}_{U|s_1,s_2}) = \delta^*(v_1,v_2|(\mathcal{P}_{\mathcal{U}}\times\mathcal{P}_{\mathcal{U}}) \cap \mathcal{A}),
%\sup_{(R_{U|s_1},R_{U|s_2})\in\mathcal{F}_{U|s_1,s_2}} \sum
\end{align}
where
\begin{equation}
    \mathcal{A} = \left\{(R_1,R_2)\in\mathbb{R}^{|\mathcal{U}|}\times\mathbb{R}^{|\mathcal{U}|} \middle| h(R_1,R_2) \leq 0\right\}, \\
\end{equation}
\begin{align}
h(R_1,R_2) &= h_1(R_1) + h_2(R_2) - C, \displaybreak[0] \\
h_1(R_1) &= \sqrt{\sum_u \frac{\hat P_{s_1,u}^2}{R_{u|s_1}}}, \displaybreak[0] \\
h_1(R_2) &= \sqrt{\sum_u \frac{\hat P_{s_2,u}^2}{R_{u|s_2}}}, \displaybreak[0] \\
C &= \sqrt{B+1} - 1 + \hat P_{s_1} + \hat P_{s_2}.
\end{align}
By Lemma \ref{lem:appint} one has
\begin{align}
&\delta^*(v_1,v_2|\mathcal{F}_{U|s_1,s_2}) \label{eq:app1} \\
&= \min_{t_1,t_2,w_1,w_2}\left\{\delta^*(t_1,t_2|\mathcal{A}) + \delta^*(w_1,w_2|\mathcal{P}_U \times \mathcal{P}_U)\right\}. \nonumber
\end{align}

By Lemma \ref{lem:appcart} we have
\begin{equation}
\delta^*(w_1,w_2|\mathcal{P}_{\mathcal{U}} \times \mathcal{P}_{\mathcal{U}}) = \delta^*(w_1|\mathcal{P}_{\mathcal{U}})+\delta^*(w_2|\mathcal{P}_\mathcal{U}).
\end{equation}
As in \eqref{eq:appf2}
\begin{align}
\delta^*(w|\mathcal{P}_U) &= \max_{P \in \mathcal{P}_{\mathcal{U}}} P^Tw \\
&= \max_u w_u.
\end{align}

Now let us consider $\delta^*(t_1,t_2|\mathcal{A})$. Applying Lemmas \ref{lem:appconst} and \ref{lem:appsum}, we get

\begin{align}
\delta^*(t_1,t_2|\mathcal{A}) = \min_{c \geq 0} \left\{h_1^*\left(\tfrac{t_1}{c}\right)+h_2^*\left(\tfrac{t_2}{c}\right)+cC\right\}, \label{eq:app2}
\end{align}
so it remains to find expressions for the $h_i^*$. This is done in Lemma \ref{lem:apphstar} below; combining this with equations \eqref{eq:app1}--\eqref{eq:app2} now proves the Theorem.
\end{IEEEproof}

\begin{lemma}
\label{lem:apphstar}
Let $h:\mathbb{R}^k\to\mathbb{R}$, $h(x)=\sqrt{\sum_{i=1}^k\frac{\kappa^2_i}{x_i}}$, with $\kappa_i>0$. Let $\lambda = \sum_{i=1}^k \kappa_i \sqrt{-v_i}$. Then
\begin{equation}
h^*(v) =
\begin{cases}
-(2^{-2/3}+2^{1/3})\lambda^{2/3},\quad &\text{if } \max_i v_i \leq 0,\\
\infty,\quad &\text{otherwise}.
\end{cases}
\end{equation}
\end{lemma}
\begin{IEEEproof}
By definition
\begin{equation}
h^*(v) = \sup_{x\in\mathbb{R}^k}\left( \sum_{i=1}^k v_ix_i - \sqrt{\sum_{i=1}^k\frac{\kappa^2_i}{x_i}}\right).
\end{equation}
Note, that if any of the $v_i$ are positive, then $h^*(v)$ is unbounded. Furthermore, for those $i\in\{1,\dots,k\}$ for which $v_i=0$, we get $x_i\to\infty$ and $0=\kappa_i\sqrt{-v_i}$ contribution to $h^*(v)$. Therefore, only need to consider $\max_i v_i < 0$ in the remainder. 
The partial derivative w.r.t.\ $x_i$ of the expression that is optimized is
\begin{equation}
v_i + \frac{1}{2\sqrt{\sum_{j=1}^k\frac{\kappa^2_j}{x_j}}}\frac{\kappa^2_i}{x_i^2}.
\end{equation}
This means that all partial derivatives are zero if $x$ is of the form
\begin{equation}
x_i = c \frac{\kappa_i}{\sqrt{-v_i}}, 
\end{equation}
for some constant $c>0$. This gives
\begin{equation} \label{eq:hstar}
h^*(v) = \sup_{c>0}\left( -c\lambda - \sqrt{\frac{\lambda}{c}} \right),
\end{equation}
where $\lambda = \sum_{i=1}^k\kappa_i\sqrt{-v_i}$.
This supremum is attained at
\begin{equation}
c = 2^{-2/3}\lambda^{-1/3}.
\end{equation}
Substituting this into \eqref{eq:hstar} completes the proof.
\end{IEEEproof}
\end{document}